\documentclass[aps,physrev,reprint,showpacs,longbibliography]{revtex4-2}
\usepackage[english]{babel}
\usepackage{amsmath,amssymb,bbm,graphicx,color,comment}
\usepackage[bookmarks=true,colorlinks,citecolor=blue,urlcolor=blue]{hyperref}
\usepackage{dsfont}
\usepackage{soul}
\usepackage{bbm}
\usepackage{float}
\usepackage{dcolumn}   
\usepackage{bm}        
\usepackage{filecontents}
\usepackage{lineno}
\usepackage{mathtools}
\usepackage[usenames,dvipsnames]{xcolor} 
\usepackage[export]{adjustbox}
\usepackage{adjustbox}
\usepackage{braket}
\usepackage{tikz}
\usepackage{bbold}
\usepackage[normalem]{ulem}
\allowdisplaybreaks[4]
\usepackage{subfigure}
\usepackage{xcolor}
\usepackage{soul}
\soulregister\cite7
\soulregister\ref7

\begin{document}
\clearpage


\title{Dissipative Quantum Battery in the Ultrastrong Coupling Regime Between Two Oscillators
}

\author{Yu-qiang Liu$^{1}$}
\email{liuyuqiang@htu.edu.cn}
\author{Yi-jia Yang$^2$}
\author{Zheng Liu$^2$}
\author{Bao-qing Guo$^3$}
\author{Ting-ting Ma$^4$}
\author{Zunlue Zhu$^1$}
\author{Wuming Liu$^{5}$}
\author{Xingdong Zhao$^1$}
\email{phyzhxd@gmail.com}
\author{Chang-shui Yu$^{2}$}
\email{ycs@dlut.edu.cn}
\affiliation{$^1$School of Physics, Henan Normal University, Xinxiang 453007, China}
\affiliation{$^2$School of Physics, Dalian University of Technology, Dalian 116024, China}
\affiliation{$^3$Quantum Information Research Center, Shangrao Normal University, Shangrao 334001, China}
\affiliation{$^4$Key Laboratory of Low-Dimensional Quantum Structures and Quantum Control of Ministry of Education,
Department of Physics and Synergetic Innovation Center for Quantum Effects and Applications, Hunan Normal University, Changsha 410081, China}
\affiliation{$^5$Beijing National Laboratory for Condensed Matter Physics, Institute of Physics, Chinese Academy of Sciences, Beijing 100190, China}

\begin{abstract}

In this work, we propose an open quantum battery that stores and releases energy by employing a two-mode ultrastrongly coupled bosonic system, with one mode (the charger) coupled to an independent heat reservoir. Our results demonstrate that both the charging energy and ergotropy of the quantum batteries can be significantly enhanced within the ultrastrong coupling regime and across a broader temperature range in transient time. A unidirectional energy flow is achieved by controlling the system’s initial state through its two-mode squeezed ground state. Furthermore, we show that the steady-state stored energy, along with its corresponding ergotropy, can be enhanced at larger temperatures and stronger coupling strengths. Notably, a purely beam-splitter or two-mode squeezing interaction yields zero ergotropy. These findings indicate that the enhanced stored energy and ergotropy of the quantum battery arises principally from the combined effects of beam-splitter and parametric amplification (squeezing) couplings. In addition, the presence of the squared electromagnetic vector potential term can prevent a phase transition and achieve a significant charging energy and high ergotropy in the deep-strong coupling regime. The results presented herein enhance our understanding of the operating principles of open bosonic quantum batteries.

\end{abstract}

\maketitle

\date{\today}

\section{Introduction}
\label{sec:intro}

The manipulation and management of energy transfer are fundamental questions in the field of quantum thermodynamics \cite{binder2018thermodynamics}, attracting significant attention. In recent years, intensive efforts have focused on small-scale devices incorporating quantum batteries for energy storage \cite{RevModPhys.96.031001}. The concept of the quantum battery was initially proposed by Alicki and Fannes \cite{PhysRevE.87.042123}. Since then, research has expanded to explore various strategies for enhancing quantum battery performance, including the exploitation of quantum phenomena such as entanglement and coherence \cite{PhysRevE.87.042123, PhysRevA.111.012204, PhysRevLett.134.010408}, repeated interactions \cite{PhysRevLett.127.100601, PhysRevResearch.5.013155}, quantum squeezing \cite{PhysRevA.108.052213}, collective effects \cite{PhysRevLett.133.243602}, and nonreciprocity \cite{PhysRevLett.132.210402}. 
Moreover, quantum battery models have been investigated across a diverse array of systems, including those based on the quantum Ising chain \cite{PRXQuantum.5.030319}, the Dicke model \cite{PhysRevB.102.245407, PhysRevLett.120.117702}, as well as two-level systems and quantum harmonic oscillators \cite{PhysRevB.98.205423}.
 At the same time, the framework of open quantum systems also provides an efficient way to enhance the potentialities of the open quantum batteries \cite{RevModPhys.96.031001}. Recently, there has been substantial interest in the charging performance of open quantum batteries based on atom and oscillator systems \cite{PhysRevB.99.035421, PhysRevLett.122.210601, PhysRevE.111.014121, PhysRevA.102.052223, d9k1-75d4, PhysRevA.102.060201, PhysRevLett.132.090401}, and Dicke and Tavis-Cummings models \cite{PhysRevA.111.022222}. So far, most proposals for quantum batteries have focused on scenarios with weak internal coupling, using the local master equation approach. Moreover, current proposals indicate that energy backflow occurs between the quantum battery and its charger during the charging process \cite{PhysRevLett.120.117702, PhysRevB.102.245407, PhysRevB.98.205423, PhysRevB.99.035421, PhysRevA.102.052223, PhysRevA.111.022222}. Hence, achieving stable, unidirectional energy flow has emerged as an important research topic.

The ultrastrong coupling between light and matter has recently attracted considerable attention \cite{RevModPhys.91.025005, WOS:000542185800009}. The so-called ultrastrong coupling $0.1 \leq {g}/\omega<1$ and deep-strong coupling regimes $g/\omega>1$ can be approached when the coupling strength of light-matter $g$ is comparable to the bare frequency $\omega$ of the bosonic mode.
Interest in these regimes has grown notably within the frameworks of cavity and circuit quantum electrodynamics. Moreover, quantum simulations \cite{RevModPhys.86.153} of the ultrastrong and deep-strong coupling regimes have been reported in superconducting circuits \cite{RevModPhys.91.025005, WOS:000416229300001, PhysRevA.95.042313} and trapped ions \cite{PhysRevA.95.063844, PhysRevResearch.6.033023, koch2023quantum}. In these regimes, several intriguing phenomena have been reexamined, including photon blockade \cite{PhysRevLett.109.193602}, detection of virtual photons \cite{PhysRevResearch.6.013008}, multiphoton quantum Rabi oscillations \cite{PhysRevA.92.063830}, critical behavior \cite{PhysRevA.86.012316}, breakdown of gauge invariance \cite{PhysRevA.98.053819}, quantum entanglement \cite{PhysRevA.111.052437} and the Purcell effect \cite{PhysRevLett.112.016401}. The ultrastrong light-matter coupling regime may thus offer alternative avenues for exploring quantum devices, such as quantum heat engines and refrigerators \cite{PhysRevE.90.052142, PhysRevE.107.044121, doi:10.1126/sciadv.adm8792}, quantum thermal diodes and transistors \cite{PhysRevE.109.064146, PhysRevE.99.032112, PhysRevLett.128.067701, PRXQuantum.5.010341}, among others. Can the ultrastrong and deep strong coupling regimes provide a stable charging mechanism?

\begin{figure}[H]
    \centering
    \includegraphics[width=0.48\textwidth]{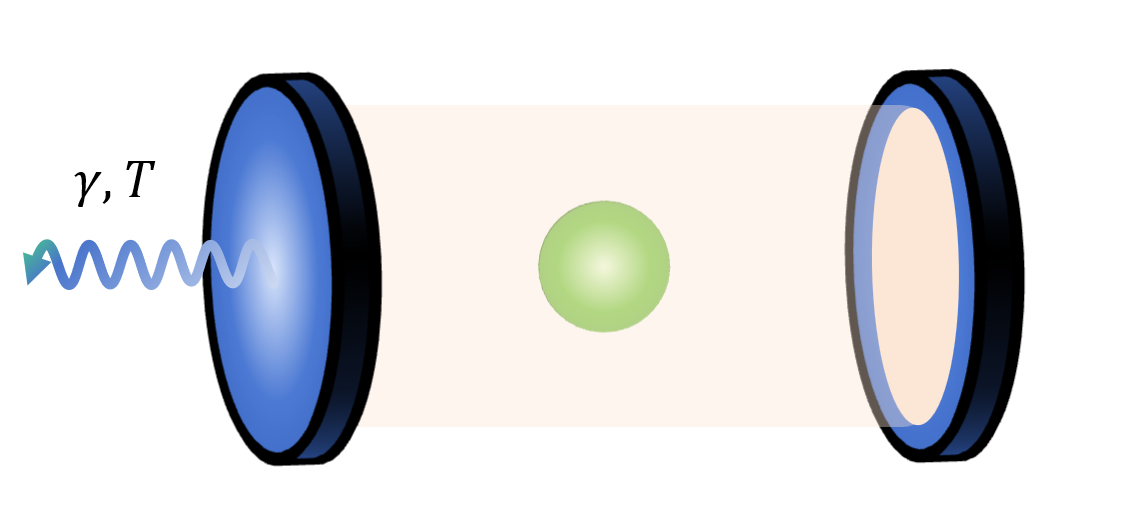}
\caption{\label{fig: model} Schematic representation of the bipartite quantum battery model. It consists of two oscillators with cavity mode $a$ (the charger) and matter $b$ mode (the battery) that are directly coupled to one another, and the mode $a$ is coupled to an independent reservoir. Here, we isolate the battery from the heat reservoir, but the charger is driven by a heat reservoir at temperature $T$.
}
\end{figure}

In this context, we propose an open quantum battery in the ultrastrong coupling regime composed of two coupled oscillators and investigate its operational mechanism using an open-system framework. Specifically, we consider a quantum battery modelled by two coupled oscillators, with one mode (the charger) interacting with an independent reservoir at finite temperature. As the ultrastrong coupling regime may present distinct characteristics compared to the weak coupling regime, we analyze the dynamics of the average populations, which characterize both the battery’s energy and the corresponding ergotropy. By controlling the initial state and the charging energy, the ergotropy can be effectively enhanced, thereby enabling unidirectional, stable charging in the ultrastrong coupling regime. Moreover, the inclusion of the squared electromagnetic vector potential $\mathbf{A}^2$ term prevents the critical point, enabling efficient charging and substantial ergotropy in the deep strong coupling regime.

The paper is organized as follows. Section \ref{Sec. II} details the physical model and its dynamics. In Section \ref{Sec. III}, we present the main results and critical observations, investigating the behavior of stored energy and ergotropy. Furthermore, to elucidate the energy storage mechanisms within the quantum battery, we analyze the roles of the beam splitter and two-mode squeezing interactions in influencing stored energy and ergotropy. Finally, we conclude in Section \ref{conclusion}. Additional technical details on the exact diagonalization procedure and related derivations are provided in Supplemental Material.

\section{The model and its dynamics} \label{Sec. II}
\subsection{Quantum battery model}

Here, we examine a model featuring two ultrastrongly coupled oscillator modes, denoted as $a$ and $b$, with mode $a$ coupled to a thermal environment at temperature $T$, as illustrated in Fig. \ref{fig: model}. The system is described by the quadratic two-mode coupled oscillators, and its Hamiltonian (in units of $\hbar=1$ and $k_{B}=1$) is given by \cite{PhysRevA.95.042313, PhysRevLett.113.093602, PhysRevLett.124.040404,PhysRevA.94.033850, baranov2020ultrastrong, PhysRevApplied.20.024039, PhysRevApplied.23.024026} 
\begin{align} \label{H_S}
H_S&=H_{0}+g(a^{\dagger} +a )(b^{\dagger} + b),\\
H_{0}&=\omega_{a} a^{\dagger} a+\omega_{b} b^{\dagger} b,
\end{align}
where $H_{0}$ is the free Hamiltonian of the oscillators, $\omega_{a}$ and $\omega_{b}$ denote their bare frequencies, and $a$, $b$ are the respective annihilation operators. 
The coupling term with $g$ represents the light-matter interaction. In the ultrastrong coupling regime, the counter-rotating terms $a^{\dagger} b^{\dagger}$ and $a b$ should be preserved \cite{PhysRevA.89.023817}.
Using the Hopfield transformation \cite{PhysRev.112.1555}, the Hamiltonian (\ref{H_S}) can be diagonalized as 
\begin{equation}
H_{S}=\omega_{+} A_{+}^{\dagger} A_{+}+\omega_{-} A_{-}^{\dagger} A_{-},
\end{equation}
where the normal frequencies are given by 
\begin{align}
\omega_{\pm}=\sqrt{\frac{\omega_a^2+\omega_b^2 \pm \omega_{c}}{2}},
\end{align}
with $\omega_c=\sqrt{(\omega_a^2-\omega_b^2)^2+16 g^2 \omega_a \omega_b}$ (see Supplemental Material). The normal operators are defined as 
\begin{align} \label{eq:A_polariton}
A_{\pm}=t_{\pm} a+u_{\pm} b+v_{\pm}  a^{\dagger}+w_{\pm} b^{\dagger}
\end{align} 
with the Hopfield coefficients provided in Supplemental Material. They satisfy the canonical commutation relations $[A_{i}, A^{\dagger}_{j}]=\delta_{i j}$, where $\delta_{i j}$ is the Kronecker delta symbol. The system becomes unstable when the interaction strength exceeds the critical coupling $g_c=\frac{\sqrt{\omega_a \omega_b}}{2}$, i.e., $g>g_{c}$. 
In the resonant case, when $g/\omega_{b}> 0.5$, the system undergoes a superradiant phase transition \cite{PhysRevE.67.066203}. It is noteworthy that the inclusion of the diamagnetic term can eliminate this critical coupling point \cite{nataf2010no, PhysRevLett.107.113602, PhysRevB.72.115303}. In the present work, we focus on the normal phase regime ($g < g_c$) to ensure the dynamical stability of the system \cite{PhysRevLett.124.040404}.

Assuming that the charger is weakly coupled to its surrounding thermal reservoir, the total Hamiltonian including the charger, battery, and reservoir, can be expressed as
\begin{align}
H_{\mathrm{tot}}=H_{S}+H_{R}+H_{R-a}.
\end{align}
The free Hamiltonian for the heat reservoir, modeled as ensembles of harmonic oscillators, is written as
\begin{equation}
H_{R}=\sum_{k}\omega_{k} {d}_{k}^{\dagger}{d}_{k},
\end{equation}
where $d^{\dagger}_{k}$, $d_{k}$ denote the creation and annihilation operators for the reservoir modes with frequencies $\omega_k$ \cite{RevModPhys.59.1}. The interaction Hamiltonian between the system and reservoir is given by
\begin{align} \label{Eq:H_SR}
H_{R-a}=\sum_k g_{k} (a^{\dagger}+a)(d^{\dagger}_{k} +d_{k}),
\end{align}
where $g_{k}$ denotes the interaction strength of system $a$ and heat reservoir with mode $k$.
By expressing the bare operators $a$ and $b$ in the normal basis as
\begin{align}  \label{original operator}
\begin{aligned}
&a=\sum_{j} (t_{j}^{*} A_{j}-v_{j} A_{j}^{\dagger}), \quad b=\sum_{j} (u_{j}^{*} A_{j}-w_{j} A_{j}^{\dagger}).
\end{aligned}
\end{align}
With the aid of Eq. (\ref{original operator}), the interaction of the system and reservoir can be rewritten as 
\begin{equation}
H_{R-a}=\sum_{j, k} g_{k} (W_{j} A_{j}^{\dagger}+W^{*}_{j} A_{j})(d_{k}^{\dagger}+d_{k}).
\end{equation} 
with $W_{j}=t_j-v_j$ for simplicity.

The master equation serves as a powerful tool for investigating the dynamics of open quantum systems. 
Within this framework, the global master equation for an open system, derived under the Born-Markov-secular approximation, takes the following form \cite{breuer2002theory}, 
\begin{align} \label{Eq:me}
\begin{split}
\frac{d \rho}{dt}&=\mathcal{L} \rho \\ &=-\mathrm{i}[H_{S}, \rho]+\sum_{j} |W_{j}|^{2} \lbrace \Gamma_{a} (-\omega_{j}) \mathcal{D}[A_{j}] \rho \\& + \Gamma_{a} (\omega_{j}) \mathcal{D}[A_{j}^{\dagger}] \rho  \rbrace,
\end{split}
\end{align} 
where $\rho$ denotes the system's density operator. The  usual Lindblad dissipator  is given by $\mathcal{D}[o] \rho=o \rho o^{\dagger}-\frac{1}{2}\left\{o^{\dagger} o, \rho\right\}$. This master equation is valid in the ultrastrong coupling regime. The steady-state solution is obtained from the null space of the superoperator $\mathcal{L}$. In our analysis, we adopt the spectral densities $\Gamma_{a}(\omega_j)=\frac{\gamma^a_{j}(\omega_{j})}{\omega_{b}} N(\omega_j)$, and $\Gamma_a (-\omega_j)=\frac{\gamma^a_{j}(\omega_{j})}{\omega_{b}}( N(\omega_j)+1)$, 
where the Bose-Einstein distribution is defined as $N (\omega_{j})=1/(\mathrm{e}^{\frac{\omega_{j}}{T_a}}-1)$. For simplicity, we consider the decay rate as $\gamma^a_{j} (\omega_{j})=\gamma^a \omega_{j}$ \cite{breuer2002theory}.

\subsection{The energy and ergotropy for the quantum battery}

The efficiency of a charging protocol for a quantum battery is evaluated based on the average energy and the ergotropy. 
Specifically, the energy in the quantum battery at $t$, $E_{b}(t)$, is determined by the expectation value of the number operator for the battery mode, $\langle b^{\dagger} b\rangle$, and is given by
\begin{equation} \label{Eb}
E_{b} (t)=\omega_{b} \langle b^{\dagger} b\rangle.
\end{equation}
The stored energy of the quantum battery is defined as
\begin{equation} \label{eq:delta_Eb}
\Delta E_{b}(t)=E_{b}(t)-E_{b}(0),
\end{equation}
where $E_{b}(t)$, and $E_{b}(0)$ represent the battery energy at time $t$ and the initial energy, respectively.
In addition, the ergotropy $\mathcal{E}$ \cite{PhysRevE.87.042123, Allahverdyan_2004}, which quantifies the maximum useful (extractable) energy stored in the quantum battery, is defined by 
\begin{equation} \label{Eq:ergotropy}
\mathcal{E} (t)=E_{b}(t)-E_{\beta}(t),
\end{equation}
where $E_{\beta}(t)=\mathrm{min}_{U_{b}} \mathrm{Tr}\lbrace U_{b} \rho_{b} U^{\dagger}_{b} H_{b}\rbrace$ with $U_{b}$ denoting an arbitrary local unitary operation on the system. For a continuous variable Gaussian system, the energy of the passive state $E_{\beta}$ can be expressed as \cite{PhysRevB.99.035421},
\begin{align} \label{Eq:M}
E_{\beta}(t)=\omega_{b}\frac{\sqrt{\mathcal{M}}-1}{2},
\end{align} 
with $\mathcal{M}=(1+2\langle b^{\dagger} b\rangle-2|\langle b\rangle|^2)^2-4|\langle b^2 \rangle-\langle b\rangle^2|^2$.

\textit{The ground-state $\rho(0)=|00 \rangle_{ab} \langle 00|$ in the bare mode basis serves as the initial state.} With this initial state, the related second moments can be solved (see Supplemental Material)
The stored energy for the quantum battery is
\begin{equation}\label{Eq:Delta_Eb_bare}
\Delta E_{b}(t)=E_{b}(t),
\end{equation}
as the initial energy of the battery is zero. With the second moments $\langle b^{\dagger} b \rangle$ and $\langle b^2 \rangle$, and zero first moments, the ergotropy of the battery can be determined.

\textit{The ground-state $|G \rangle=|0 \rangle_{+} |0 \rangle_{-}$ in the normal mode basis serves as the initial state.} In this case, $\langle A_{j} A_{j}^{\dagger}\rangle|_{t=0}=1$, while other second moments are zero. For the initial state $|G \rangle$, the relevant quantities can be solved as 
\begin{align} \label{eq:second_moments_p}
\begin{split}
\langle A_{j}^{\dagger} A_{j} \rangle&=(1-\mathrm{e}^{-\gamma^{a}_{j} |W_j|^2 t}) N^{a}(\omega_j),\\
\langle A_{j} A_{j}^{\dagger} \rangle&=1+(1-\mathrm{e}^{-\gamma^{a}_{j} |W_j|^2 t}) N^{a}(\omega_j),
\end{split}
\end{align}
with other quantities being zero. In the long time limit, $\langle A^{\dagger}_{j} A_{j}\rangle$ can be solved as
$\lim _{t \rightarrow \infty} \langle A^{\dagger}_{j} A_{j}\rangle =N^{a} (\omega_j)$.
With Eq. (\ref{original operator}) and the Hopfield coefficients, the average occupation of mode $b$ can be solved as
\begin{align}
\left\langle b^{\dagger} b \right\rangle=\sum_{j}[(|u_{j}|^{2}+ |w_{j}|^{2})\langle A_{j}^{\dagger} A_{j}\rangle+|w_j|^2].
\end{align}
Hence, the energy in the battery for time $t$ can be expressed as
$E_{b}(t)=\omega_{b} \sum_{j}[(|u_{j}|^{2}+ |w_{j}|^{2})\langle A_{j}^{\dagger} A_{j}\rangle+|w_j|^2]$.
The initial energy of the quantum battery can be written as 
\begin{align}
E_{b}(0)=\omega_{b} \sum_{j} |w_{j}|^2.
\end{align}
The stored energy of the quantum battery is
\begin{equation} \label{Eq.qb_energy_bias}
\Delta E_{b}(t)=\omega_{b} \sum_{j} (|u_{j}|^{2}+ |w_{j}|^{2})\langle A_{j}^{\dagger} A_{j}\rangle.
\end{equation}
In this case, one can express $\mathcal{M}$ as $\mathcal{M}=(1+2 \langle b^{\dagger} b\rangle)^2-4 |\langle b^2 \rangle|^2$ since the first moments are zero with $ \langle b \rangle=0$. Hence $\mathcal{M}=(1+2 \sum_{j}[(|u_{j}|^{2}+ |w_{j}|^{2})\langle A_{j}^{\dagger} A_{j}\rangle+|w_j|^2])^2-4 |-\sum_{j} u_{j}^{*} w_{j}(2 \langle A^{\dagger}_{j} A_{j}\rangle +1)|^2$. The corresponding ergotropy can be written as 
$\mathcal{E}=\omega_{b}\lbrace \sum_{j}[(|u_{j}|^{2}+ |w_{j}|^{2})\langle A_{j}^{\dagger} A_{j}\rangle+|w_j|^2]-\frac{1}{2}[(1+2 \sum_{j}(|u_{j}|^{2}+ |w_{j}|^{2})\langle A_{j}^{\dagger} A_{j}\rangle+|w_j|^2])^2 - 4 |-\sum_{j} u_{j}^{*} w_{j} (2 \langle A_{j}^{\dagger} A_{j} \rangle+1)|^2]^{\frac{1}{2}}+\frac{1}{2}\rbrace$.

\begin{figure}[H]
    \centering
\includegraphics[width=1\columnwidth]{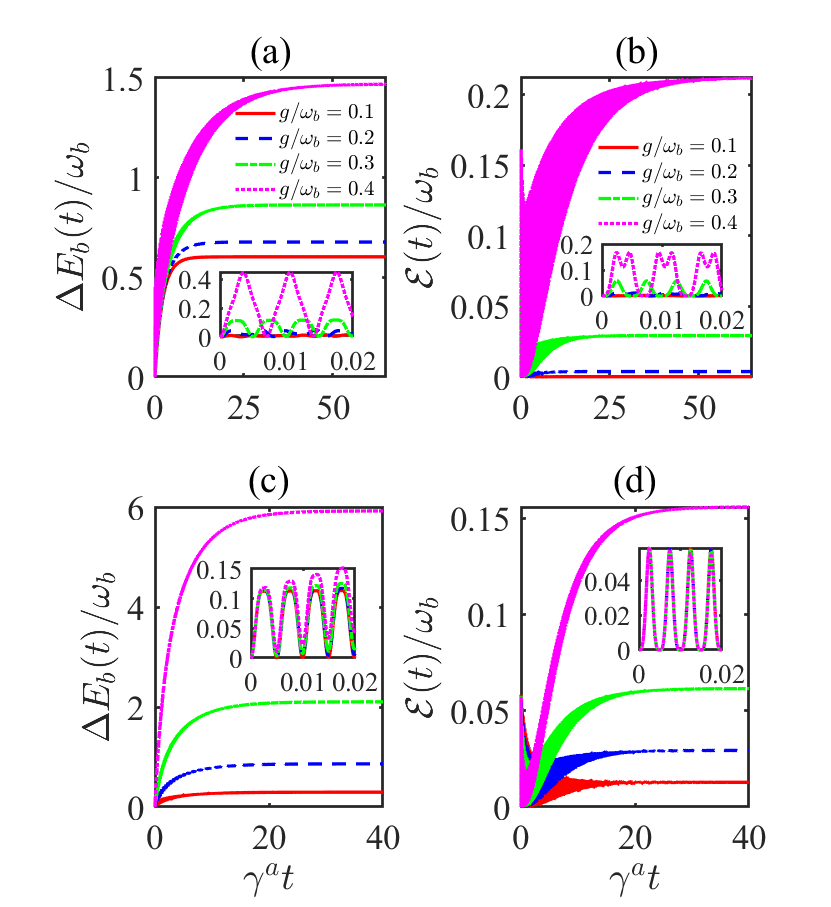}
\caption{\label{fig:qb0}
The behavior of the stored energy, $\Delta E_{b}$, and its corresponding ergotropy, $\mathcal{E}$ are presented as a function of time $t$ for various coupling strengths in panels (a) and (b), and different temperatures in panels (c) and (d) under resonant condition when considering the initial state $\rho(0)=|00 \rangle_{ab} \langle 00|$. In panels (c) and (d), the red solid, blue dashed, green dashed-dotted, and magenta dotted lines correspond to temperatures with $T_a/\omega_b=0.5, 1, 2, 5$. The following parameters are used: (a) and (b) $T_{a}=\omega_{b}$; (c) and (d) $g=0.3 \omega_b$. Other parameters are $\omega_{a}=\omega_{b}=\omega$, and $\gamma^{a}=10^{-3} \omega_{b}$. All parameters are expressed in units of the frequency $\omega_b$.}
\end{figure}

\section{Result and discussion} \label{Sec. III}

Here, we examine the behavior of quantum batteries in the ultrastrong coupling regime. We focus on the stored energy and ergotropy of the quantum batteries under the resonant conditions, where $\omega_{a}=\omega_b=\omega$. Initially, we consider the initial state $\rho(0)=|00 \rangle_{ab} \langle 00|$. In this case, according to Eq. (\ref{Eq:Delta_Eb_bare}), the quantum battery starts with zero energy, and both the energy and ergotropy exhibit Rabi-like oscillations over time before gradually increasing to reach a steady state, as shown in Fig. \ref{fig:qb0}. In the long-time limit, the maximum stored energy can be solved as 
\begin{align}
\begin{split}
\lim _{t \rightarrow \infty} \Delta E_{b}&=\omega_b\lbrace \mathcal{N}_{-}^2[1+N(\omega_-)(1+\frac{(-g+\omega+\omega_-)^2}{g^2})]\\&+ \mathcal{N}_{+}^2[1+N(\omega_{+})(1+\frac{(g+\omega+\omega_+)^2}{g^2})]\rbrace.
\end{split}
\end{align}
From this expression, we find that the steady-state stored energy is independent of the dissipation strength, facilitating stable charging. Moreover, increasing the coupling strength $g$ can enhance the stored energy in the quantum battery. It should be noted that, because the two modes are resonant, only the energy stored in the battery is depicted, as the charger contains an equivalent amount of energy. The oscillatory behavior primarily arises from the coherence terms, such as $\langle A_{j}^2 \rangle$ and $\langle A_{-}^{\dagger} A_{+} \rangle$ in the second moments, as indicated in Eq. (S7) (see Supplemental Material). Furthermore, Fig. \ref{fig:qb0} (c) and Fig. \ref{fig:qb0} (d) demonstrate that higher temperatures enable more efficient energy storage and ergotropy in the quantum battery.

\begin{figure}[H]
    \centering
\includegraphics[width=1\columnwidth]{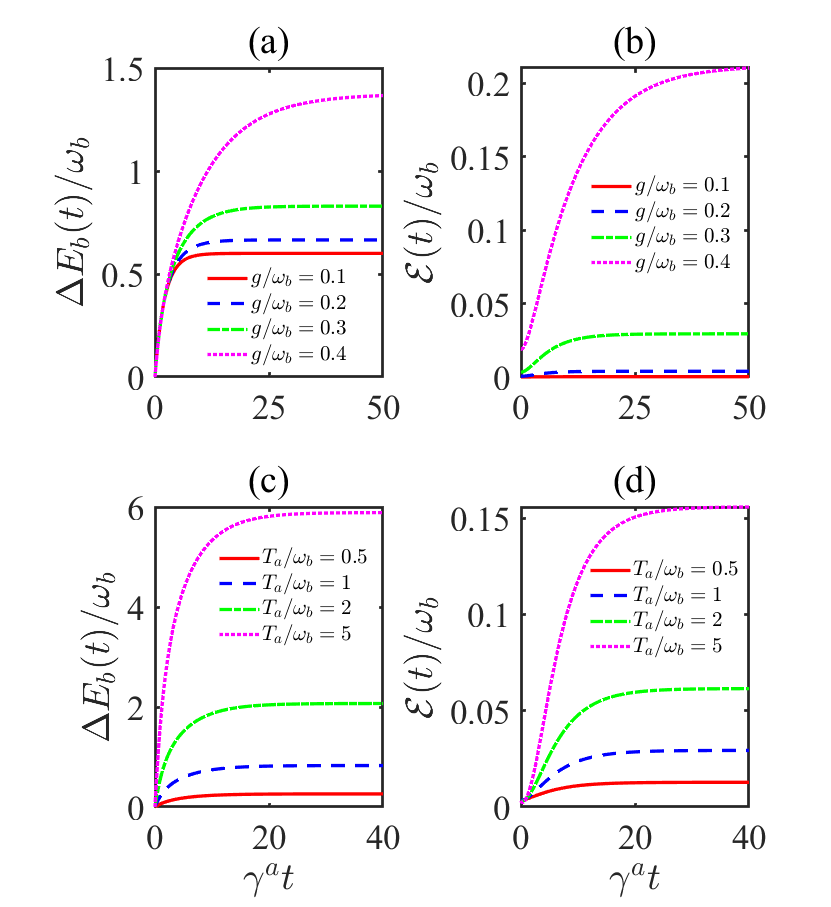}
\caption{\label{fig:qb1}
The behavior of the stored energy $\Delta E_{b}$, and its corresponding ergotropy $\mathcal{E}$ are presented as a function of time $t$ for various coupling strengths in panels (a) and (b), and different temperatures in panels (c) and (d) under resonant condition considering the initial state $|G \rangle=|0 \rangle_{+} |0 \rangle_{-}$. The following parameters are used: (a) and (b) $T_{a}=\omega_{b}$; (c) and (d) $g=0.3 \omega_b$. Other parameters are $\omega_{a}=\omega_{b}=\omega$, and $\gamma^{a}=10^{-3} \omega_{b}$. All parameters are expressed in units of the frequency $\omega_b$.}
\end{figure}

To achieve stable charging, one should let population terms $\langle A_{j}^{\dagger} A_{j}\rangle$ survive and all coherence terms, for example, $\langle A_{+}^{\dagger} A_{-}\rangle$ disappear by taking the particular initial state since the population and coherence terms evolve independently as shown in Eq. (S3) (refer to Supplemental Material). Hence we consider the ground state $|G \rangle=|0 \rangle_{+} |0 \rangle_{-}$ in the normal representation as the initial state. The stored energy of the quantum battery is given by
\begin{align}
E_{b}(t)=\omega_{b} (\bar{N}_{+}(t)+\bar{N}_{-}(t)),
\end{align}
where $\bar{N}_{\pm}=\mathcal{N}_{\pm}^2 [1+(1-\mathrm{e}^{-\gamma_{\pm}^{a} |W_{\pm}|^2 t}) N(\omega_{\pm})(1+(\pm 1+\frac{\omega+\omega_{\pm}}{g})^2)]$ together with the normalized coefficients $\mathcal{N}_{\pm}=\frac{g}{\sqrt{2 (\omega+\omega_{\pm})(\pm 2 g+\omega+\omega_{\pm})}}$. 
Accordingly, the stored energy is expressed as
 \begin{equation}
\Delta E_{b}(t)=\omega_{b}(\bar{N}_{+}(t)+\bar{N}_{-}(t)-\mathcal{N}_{+}^2-\mathcal{N}_{-}^2).
\end{equation} 

\begin{figure}[H]
    \centering
\includegraphics[width=0.7\columnwidth]{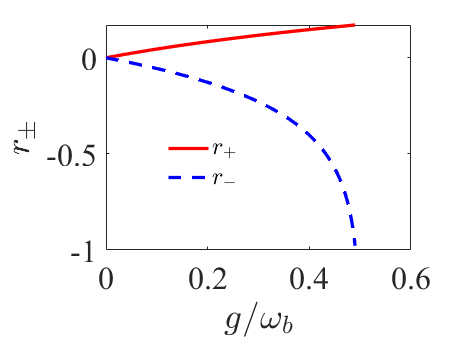}
\caption{\label{fig:squeezing}
The squeezing parameters $r_{\pm}$ are plotted as a function of the coupling strength $g$. The parameter can take $\omega_{a}=\omega_{b}=\omega$.}
\end{figure}

As illustrated in Fig. \ref{fig:qb1} (a), the stored energy of the quantum battery monotonically increases from charger to battery with time during the transient time until it asymptotically reaches a steady state. It means that the unidirectional energy transfer can be realized. Moreover, the stored energy can be enhanced as the coupling strength increases. The steady-state charging scenario is independent of initial states, as stated before. In the long-time limit, the stored energy reaches its maximum value, which is given by
\begin{align}
\begin{split}
\lim _{t \rightarrow \infty} \Delta E_{b}&=\omega_b[ \mathcal{N}_{-}^2 N(\omega_{-})(g^2+(-g+\omega+\omega_{-})^2)\\&+\mathcal{N}_{+}^2 N(\omega_{+}) (g^2+(g+\omega+\omega_{+})^2)]/g^2.
\end{split}
\end{align}
The unidirectional energy transfer originates from the survival of population terms $\langle A_{j}^{\dagger} A_{j}\rangle$ and disappearance of coherence terms such as $\langle A_{j}^2 \rangle$ and $\langle A_{-}^{\dagger} A_{+} \rangle$ by the ground state $G=|0\rangle_{+} |0\rangle_{-}$ for two ultrastrong coupling oscillators in the normal mode representation as shown in Eqs. (\ref{eq:second_moments_p}) and (S3) (refer to Supplemental Material). This mechanism is distinct from that observed in systems of weakly coupled oscillators described by the local master equation, as demonstrated in Refs. \cite{PhysRevB.99.035421, PhysRevLett.132.210402}. In the weak coupling regime $g \sim \gamma$, the terms of population and coherence are mutually dependent. Moreover, the coherence terms in the normal mode basis persist (refer to Eq. (S16) in Supplemental Material) and unidirectional energy transfer occurs only in the overdamped regime \cite{PhysRevB.99.035421}, as detailed in Supplemental Material. The unidirectional energy transfer can also be attained through nonreciprocal coupling by introducing a shared reservoir, as discussed in Ref. \cite{PhysRevLett.132.210402}. Therefore, unidirectional and stable charging can be achieved through diverse physical mechanisms. According to Eq. (\ref{Eq:ergotropy}), the ergotropy can be obtained, and its evolution over time for different coupling strengths is depicted in Fig. \ref{fig:qb1} (b). It is found that the ergotropy increases monotonically with time until it approaches the steady state. To determine the steady-state ergotropy, we can solve $\mathcal{M}$ as $\mathcal{M}=[g^2 (1+\mathcal{N}_{-}^2(4+8 N(\omega_{-}))+\mathcal{N}_{+}^2(4+8 N(\omega_{+})))-2 g \mathcal{N}_{-}^2 (1+4 N (\omega_{-}))(\omega+\omega_{-})+2 g \mathcal{N}_{+}^2 (1+4 N(\omega_{+}))(\omega+\omega_{+})+2(\mathcal{N}_{-}^2 N(\omega_{-})(\omega+\omega_{-})^2+\mathcal{N}_{+}^2 N(\omega_{+})(\omega+\omega_{+})^2)] [g^2+2 g \mathcal{N}_{-}^2(\omega+\omega_{-})-2 g \mathcal{N}_{+}^2 (\omega+\omega_{+})+2(\mathcal{N}_{-}^2 N(\omega_{-})(\omega+\omega_{-})^2+\mathcal{N}_{+}^2 N(\omega_{+})(\omega+\omega_{+})^2)]/g^4$. We also find that the ergotropy can be enhanced by increasing the temperature, as shown in Fig. \ref{fig:qb1} (c) and Fig. \ref{fig:qb1} (d). Consequently, unidirectional and stable energy flow from the heat reservoir to the battery is achieved via the charger, in contrast to the closed system scenario described in Supplemental Material.

\begin{figure}[H]
    \centering
\includegraphics[width=1\columnwidth]{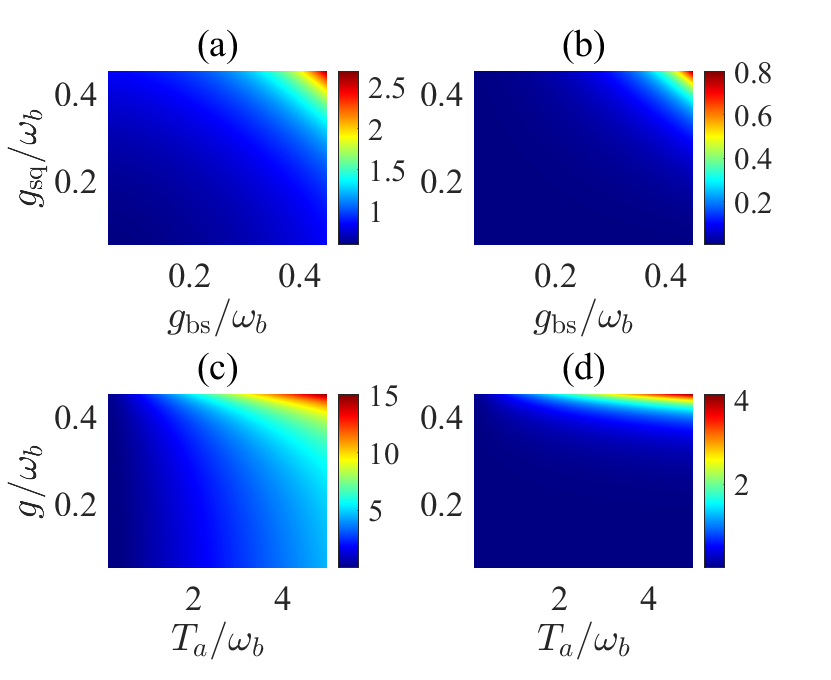}
\caption{\label{fig:qb2}
Behavior of the stored energy $\Delta E_{b}$ and corresponding ergotropy $\mathcal{E}$ as functions of the coupling strengths $g_{\mathrm{bs}}$ and $g_{\mathrm{sq}}$ (a, b), and temperature $T_a$ and coupling strength $g$  (c, d) under resonant condition. The parameters are set as $T_a=\omega_b$ for panels (a) and (b), $g_{\mathrm{bs}}=g_{\mathrm{sq}}=g=0.3 \omega_{b}$ for panels (c) and (d), and other parameters can take $\omega_{a}=\omega_{b}=\omega$, with all parameters expressed in units of the frequency $\omega_b$.}
\end{figure}

To elucidate the squeezing properties of the system, we recast the Hamiltonian (\ref{H_S}) as $H= (\omega-g) c^{\dagger} c+(\omega+g) d^{\dagger} d-\frac{g}{2} (c^2+c^{\dagger 2})+\frac{g}{2}(d^2+d^{\dagger 2})$ at resonance by introducing the two hybrid modes given by $c=\frac{a-b}{\sqrt{2}}$, $d=\frac{a+b}{\sqrt{2}}$ \cite{PhysRevA.95.042313}. Subsequently, two distinct squeezing transformations are applied to further diagonalize the Hamiltonian. For the resonant case, the ground state of the system is determined as $| G \rangle =\mathrm{e}^{\frac{r_{-} (c^2-c^{\dagger 2})}{2}}\mathrm{e}^{\frac{r_{+} (d^2-d^{\dagger 2})}{2}} |0 \rangle_{a} |0 \rangle_{b}$, where $r_{\pm}=\frac{1}{4} \log(1\pm\frac{2 g}{\omega})$. The corresponding eigenfrequencies are then given by $\omega_{\pm}=\omega\sqrt{(1\pm\frac{2 g}{\omega})}=\omega \mathrm{e}^{2 r_{\pm}}$. It is obviously that the squeezing coefficient becomes $r \rightarrow 0$ when $\frac{g}{\omega} \rightarrow 0$. In the limit $\frac{g}{\omega} \ll 1$, the rotating wave approximation holds, and $r_{\pm}\simeq \frac{1}{4} (\pm \frac{2 g}{\omega})\ll 1$. Therefore, the squeezing effects are negligible, implying that the contribution from the two-mode squeezing interaction to the overall squeezing property is minimal. The related discussion can refer to Refs. \cite{cxvs-5pb1, PhysRevA.110.063703, PhysRevA.85.043821}. As illustrated in Fig. \ref{fig:squeezing}, the squeezing parameters increase with the coupling strength $g$. This observation indicates that the system's ground state exhibits quantum correlations \cite{PhysRevB.72.115303}, and that the squeezing facilitates energy charging. Accordingly, we can employ this two-mode squeezed state as the initial state to realize unidirectional energy flow, as previously analyzed.

\begin{figure}[H]
    \centering
\includegraphics[width=1\columnwidth]{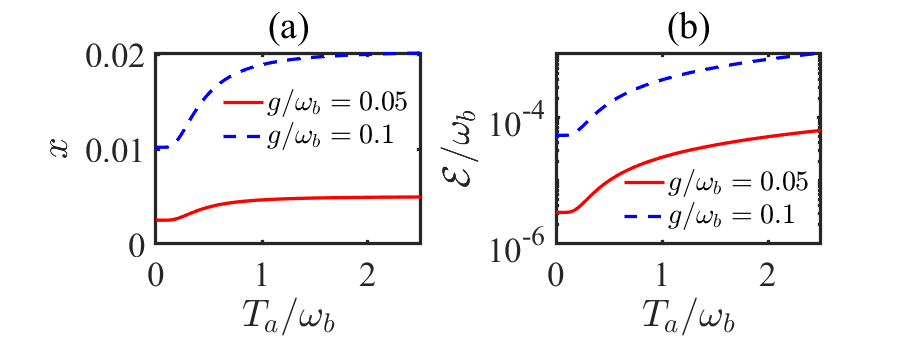}
\caption{\label{fig1_}
The $x$ (a) and ergotropy (b) as functions of temperature for different coupling strengths. The parameter can take $\omega_{a}=\omega_{b}=\omega$, with all parameters expressed in units of the frequency $\omega_b$.}
\end{figure}

To optimize the quantum battery's performance, one can rewrite the interaction Hamiltonian of the charger and battery as
\begin{equation} \label{Eq:full_Hamilton}
H_{I}=g_{\mathrm{bs}} (a^{\dagger} b+a b^{\dagger})+g_{\mathrm{sq}} (a^{\dagger} b^{\dagger}+ a b),
\end{equation}
where the terms proportional to $g_{\mathrm{bs}}$ and $g_{\mathrm{sq}}$ represent the beam-splitter and two-mode squeezing interactions, respectively \cite{RevModPhys.84.621, RevModPhys.86.1391}. The anisotropy coupling strength with $g_{\mathrm{bs}} \neq g_{\mathrm{sq}}$ can be realized in magnon–magnon coupling system \cite{makihara2021ultrastrong}. As shown in Fig. \ref{fig:qb2} (a) and Fig. \ref{fig:qb2} (b), the cooperative effect of the beam-splitter and two-mode squeezing interactions can realize the larger energy and ergotropy. Let us consider the impact of the heat-reservoir temperature $T_{a}$ and coupling strength $g_{\mathrm{bs}}=g_{\mathrm{sq}}=g$ on the energy and ergotropy of the steady-state quantum battery, as illustrated in Fig. \ref{fig:qb2} (c) and Fig. \ref{fig:qb2} (d). Fig. \ref{fig:qb2} (c) and Fig. \ref{fig:qb2} (d) indicate that the steady-state values of stored energy and ergotropy increase with temperature $T_a$ and coupling strength $g$. This occurs because, under higher temperatures and stronger coupling strengths, the battery's state deviates obviously from a thermal (passive) state. It reflects that the higher-temperature reservoir provides greater energy, which is subsequently transferred to the battery via the charger. From Eqs. (\ref{Eq:ergotropy}) and (\ref{Eq:M}), the ergotropy is given as 
$\mathcal{E}=E_{b}-E_{\beta}=\omega_{b}[\langle b^{\dagger} b\rangle-\frac{\sqrt{\mathcal{M}}-1}{2}]$,
with $\mathcal{M}=(1+2\langle b^{\dagger} b\rangle)^2-4|\langle b^2 \rangle|^2$,
under the condition that $\langle b \rangle=0$. Notably, this quantity exhibits a nonlinear dependence on the field operator second moments. In particular, the ergotropy vanishes when $\langle b^2 \rangle = 0$. Thus, $\langle b^2 \rangle$ serves as a primary indicator of whether the battery possesses any ergotropy. Moreover, $\langle b^2 \rangle$ originates from the ultrastrong coupling of two oscillators \cite{PhysRevB.72.115303}. When the coupling strength $g$ and temperatures $T_a$ are not large, the following approximation holds,
\begin{align}
\begin{split}
\sqrt{\mathcal{M}}&=\sqrt{(1+2\langle b^{\dagger} b\rangle)^2-4|\langle b^2 \rangle|^2}\\&
=\sqrt{(1+2\langle b^{\dagger} b\rangle)^2[1-\frac{4|\langle b^2 \rangle|^2}{(1+2\langle b^{\dagger} b\rangle)^2}]} \\& 
\simeq (1+2\langle b^{\dagger} b\rangle)[1-\frac{2 |\langle b^2 \rangle|^2}{(1+2\langle b^{\dagger} b \rangle)^2}],
\end{split}
\end{align}
where we let $x=\frac{2|\langle b^2\rangle|}{1+2\langle b^{\dagger} b \rangle}$ and use the approximation $\sqrt{1-x^2} \simeq 1-\frac{x^2}{2}$ for small $x$. As shown in Fig. \ref{fig1_} (a), this approximation is valid. The ergotropy can be approximated by
$\mathcal{E}=\frac{2|\langle b^2\rangle|^2}{1+2\langle b^{\dagger} b \rangle}$. Moreover, as illustrated in Fig. \ref{fig1_} (b), the ergotropy increases with temperature.

\begin{figure}[H]
    \centering
\includegraphics[width=1\columnwidth]{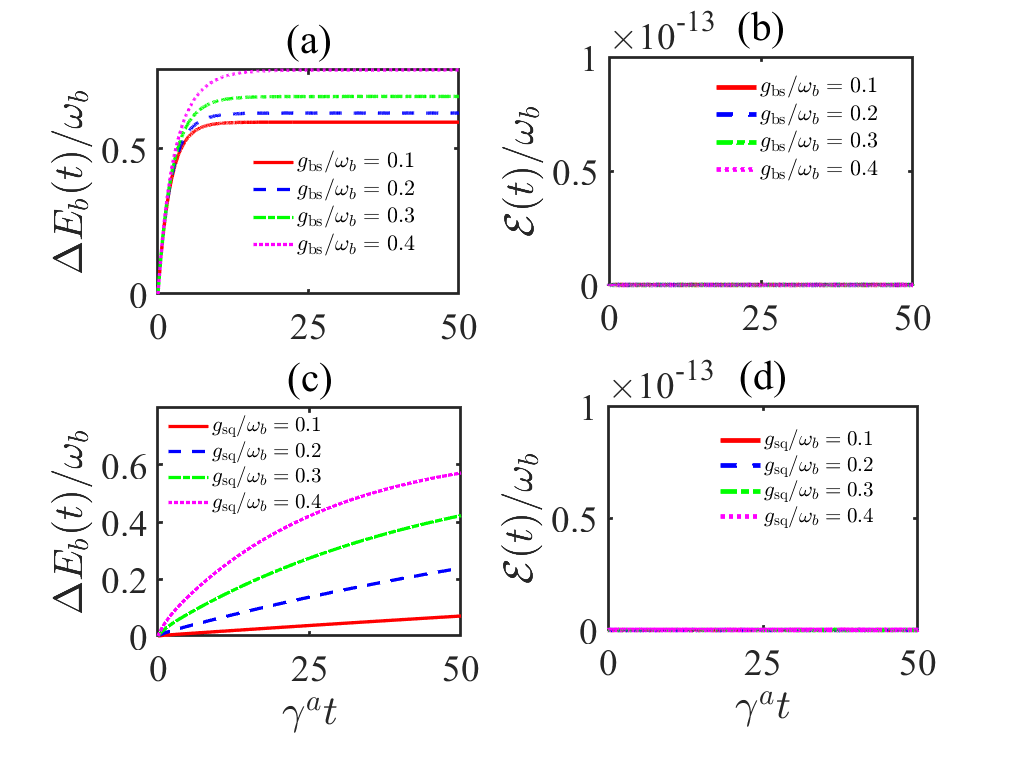}
\caption{\label{fig:qb3}
Behavior of the stored energy $\Delta E_{b}$ (a, c) and ergotropy $\mathcal{E}$ (b, d) as a function of the time $t$ for different coupling strengths $g_{\mathrm{bs}}$ and $g_{\mathrm{sq}}$ at resonance based on initial state $|G \rangle=|0 \rangle_{+} |0 \rangle_{-}$. In panels (a) and (b), the squeezing coupling parameter is set to $g_{\mathrm{sq}}=0$, while in panels (c) and (d), the beam-splitter coupling parameter is set to $g_{\mathrm{bs}}=0$. Other parameters are $\omega_{a}=\omega_{b}=\omega$, $\gamma^{a}=10^{-3} \omega_{b}$ and $T_{a}=\omega_{b}$, with all quantities expressed in units of the frequency $\omega_{b}$.}
\end{figure}

To understand these results, according to Eqs. (S2) in Supplemental Material and (\ref{Eq.qb_energy_bias}), one can see that the stored energy of the quantum battery is
\begin{align}
\begin{split}
\Delta E_{b}(t)&=\omega_{b} \sum_{j} N_{j}^2 \delta_{j}^2 N(\omega_{j})(1-\mathrm{e}^{-\gamma_{j}^{a} |W_{j}|^2 t}),
\end{split}
\end{align}
where $N_{\pm}=g_{\mathrm{sq}}/\sqrt{-2 g_{\mathrm{sq}}^2+2(\pm g_{\mathrm{bs}}+\omega+\omega_{\pm})^2}$ are the normalized coefficients and $\delta_{\pm}=1+\frac{\pm g_{\mathrm{bs}}^2 \omega+ g_{\mathrm{bs}} g_{\mathrm{sq}} (\omega+\omega_{\pm})}{g_{\mathrm{bs}} g_{\mathrm{sq}} \omega}$. The corresponding ergotropy can be solved by employing
$\mathcal{M}=(1+2(N_{-}^2+N_{+}^2+(1-\mathrm{e}^{-\gamma_{-}^{a} |W_{-}|^2 t})N(\omega_{-}) N_{-}^2(1+\xi_{-}^2)+(1-\mathrm{e}^{-\gamma_{+}^{a} |W_{+}|^2 t} N(\omega_{+}) N_{+}^2 (1+\xi_{+}^2)))^2- 4|-N_{-}^2(1+2(1-\mathrm{e}^{-\gamma_{-}^{a} |W_{-}|^2 t}) N(\omega_{-}))\xi_{-}+N_{+}^2(1+2(1-\mathrm{e}^{-\gamma_{+}^{a} |W_{+}|^2 t}) N(\omega_{+}))\xi_{+}|^2$
with $\xi_{\pm}=\frac{(\pm g_{\mathrm{bs}}+\omega+\omega_{\pm})^2}{g_{\mathrm{sq}}^2}$. Here we also discuss the scenario in which the initial state is given by $|G \rangle=|0 \rangle_{+} |0 \rangle_{-}$. As shown in Fig. \ref{fig:qb3} (a) and Fig. \ref{fig:qb3} (c), unidirectional energy storage within the quantum battery is achievable using beam-splitter or two-mode squeezing interactions. A comparison between Fig. \ref{fig:qb3} (b) and Fig. \ref{fig:qb3} (d) indicates that no ergotropy is produced under the purely beam-splitter or two squeezing interaction.
Notably, according to Eq. (\ref{Eq:ergotropy}), $\mathcal{E}=\omega_{b} (\langle b^{\dagger} b \rangle-\frac{\sqrt{\mathcal{M}}-1}{2})$ and $\mathcal{M}=(1+2 \langle b^{\dagger} b \rangle)^2$, the ergotropy is obtained as $\mathcal{E}_{b}$=0. Overall, the foregoing discussion demonstrates that ergotropy $\mathcal{E}$ primarily arises from the combined effects of the beam splitter and squeezing interactions.

\begin{figure}[H]
    \centering
\includegraphics[width=1\columnwidth]{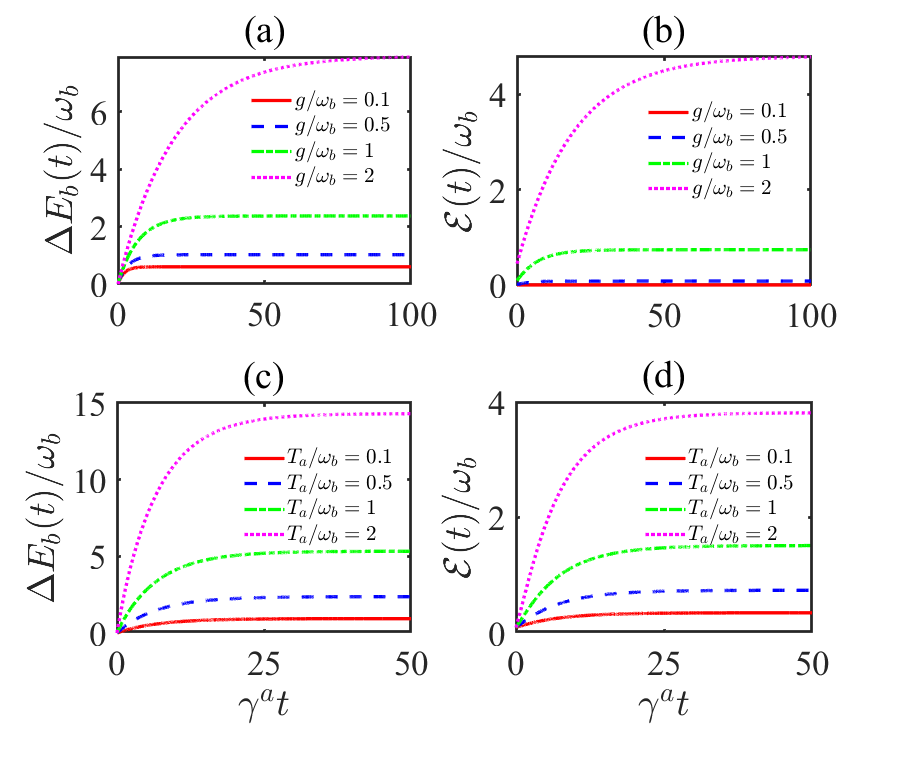}
\caption{\label{fig:qb_A2}
Behavior of the stored energy $\Delta E_{b}$ for the Hopfield model as shown in Eq. (\ref{eq:Hop}), depicted in panels (a) and (c), and its corresponding ergotropy $\mathcal{E}$ shown in panels (b) and (d), are plotted as functions of time $t$ for various coupling strengths $g$ and temperatures $T_a$. In panels (a) and (b), the temperature parameter is set to $T_{a}=\omega_b$, and in panels (c) and (d), the coupling strength is fixed at $g/\omega_{b}=1$. The remaining parameters are specified as $\omega_{a}=\omega_{b}=\omega$, $\gamma^{a}=10^{-3} \omega_{b}$, with all values expressed in units of the frequency $\omega_b$.}
\end{figure}

\begin{figure}[H]
    \centering
\includegraphics[width=1\columnwidth]{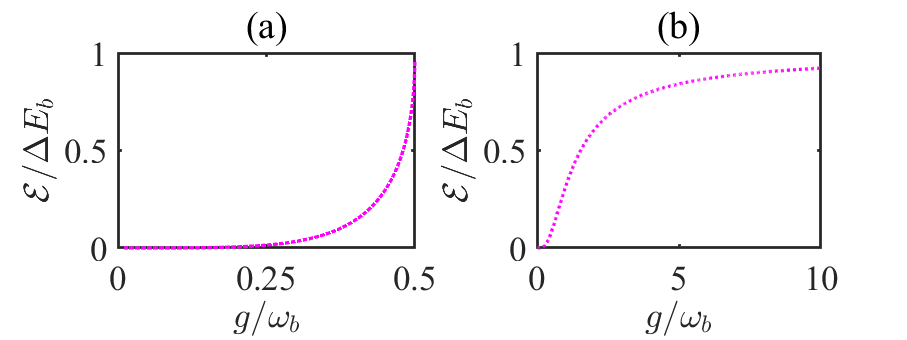}
\caption{\label{fig:ratio}
The ratio of the ergotropy to the stored energy $\mathcal{E}/\Delta E_{b}$ as a function of the coupling strength $g$ for two coupled oscillators without $\mathbf{A}^2$ term (a) and the Hopfield model (refer to Eq. (\ref{eq:Hop})) (b). The parameter can take $\omega_{a}=\omega_{b}=\omega$ and $T_a/\omega_b=1$.}
\end{figure}

Further, we investigate the influence of the squared electromagnetic vector potential $\mathbf{A}^2$ term with $D(a+a^{\dagger})^2$ 
\cite{ PhysRevLett.35.432} on the energy and ergotropy of the quantum battery. The presence of $D(a+a^{\dagger})^2$ in the Dicke Hamiltonian can eliminate the quantum critical point and inhibit the superradiant phase transition \cite{PhysRevA.86.063831} under the Thomas-Reiche-Kuhn (TRK) sum rule \cite{nataf2010no, PhysRevLett.107.113602}.  The Hamiltonian based on Eq. (\ref{H_S}) for Hopfield model is expressed as \cite{PhysRevLett.112.016401, PhysRevA.91.063840}
\begin{align} \label{eq:Hop}
H_{\mathrm{Hop}}=H_{S}+D (a+a^{\dagger})^2.
\end{align} 
In the simulation, we set $D=g^2/\omega_{b}$. 
Under these conditions, the lower normal frequency $\omega_{-}$ approaches zero as the coupling strength satisfies $g/\omega \rightarrow \infty$. 
Based on Eq. (\ref{Eq.qb_energy_bias}), the stored energy is
\begin{align}
\begin{split}
\Delta E_{b}(t)&=\Omega_{-} N(\omega_{-}) \cos^2 \theta (1-\mathrm{e}^{-t \gamma^{a}_{-} |W_{-}|^2}) \\&+\Omega_{+}N(\omega_{+})\sin^2\theta (1-\mathrm{e}^{-t \gamma^{a}_{+} |W_{+}|^2}) ,
\end{split}
\end{align}
where $\Omega_{\pm}=\frac{\omega_b^2+\omega_{\pm}^2}{\omega_{\pm}}$, and
$\omega_{\pm}=\sqrt{\frac{\omega_{m}^2+\omega_b^2}{2} \pm \sqrt{\left(\frac{\omega_{m}^2-\omega_b^2}{2}\right)^2+4 g^2 \omega_a \omega_b}}$ are the normal frequencies with $\omega_{m}^2=\omega_a^2+4 D \omega_a$ and $\sin 2 \theta=-\frac{4 g \sqrt{\omega_a \omega_b}}{\omega_{+}^2-\omega_{-}^2}$ \cite{PhysRevA.91.063840}. The steady-state energy of the quantum battery $E_b$ is obtained as $E_b=\frac{1}{4} (-2+[\frac{(1+2 N(\omega_{-})) \cos^2 \theta (\omega_{b}^2+\omega_{-}^2)}{ \omega_{-}}+\frac{(1+2 N(\omega_{+}))\sin^2\theta(\omega_{b}^2+\omega_{+}^2)}{ \omega_{+}}])$, and the stored energy is given by 
\begin{align}
\Delta E_{b}= \Omega_{-} N(\omega_{-}) \cos^2 \theta +\Omega_{+} N(\omega_{+})\sin^2\theta.
\end{align}
The corresponding ergotropy can be solved using the expression $\mathcal{M}$, as shown in Supplemental Material. 
Fig. \ref{fig:qb_A2} displays the energy and ergotropy of the quantum battery as functions of time for varying coupling strengths $g$ and temperatures $T_a$. Notably, both the energy and ergotropy are significantly enhanced, for example, in the deep strong coupling regime $g/\omega_{b}=2$ or at higher temperatures $T_{a}/\omega_{b}=2$ in the presence of $\mathbf{A}^2$ term. In addition, we analyze an alternative performance index for the battery in both models, as shown in Eqs. (\ref{H_S}) and (\ref{eq:Hop}), which is characterized by the ratio of the ergotropy to the stored energy. As illustrated in Fig. \ref{fig:ratio} (a), a stronger coupling can enhance the performance of the battery for the two ultrastrongly coupled oscillators without the $\mathbf{A}^2$ term. Similarly, the deep strong coupling regime in the Hopfield model appears to improve performance, as shown in Fig. \ref{fig:ratio} (b).

Besides, coupled oscillators systems can be implemented using current solid-state technologies, such as superconducting quantum circuits \cite{PhysRevA.95.042313, PhysRevLett.113.093602, PhysRevLett.124.040404, PhysRevApplied.23.024026}, interacting nanomechanical oscillators \cite{RevModPhys.86.1391}, and coupled systems of spin-wave excitations (magnons) and microwave photons \cite{PhysRevApplied.20.024039}. Moreover, a general noiseless resistor paired with an associated fluctuating voltage source can be employed to model a heat reservoir  \cite{RevModPhys.82.1155}. For instance, the parameters may be chosen such that $\omega_b/2\pi=6 \mathrm{GHz}$ for superconducting circuits \cite{PhysRevA.95.042313} and $\omega_a/2\pi=8.65 \mathrm{GHz}$ for the coupled magnon-photon system \cite{PhysRevApplied.20.024039}, with other parameters normalized by the respective frequency.

\section{Conclusion}
\label{conclusion}

To summarize, we introduce a quantum battery and investigate the use of a heat reservoir to power a charger coupled to the battery via a system of two ultrastrongly coupled oscillators. Our results indicate that the energy storage in the quantum battery and its corresponding ergotropy are enhanced with increasing coupling strength and elevated reservoir temperatures. Unidirectional energy charging can be achieved by modulating the system's initial state. A purely beam-splitter or two-mode squeezing interaction fails to produce ergotropy. Overall, the combined effect of the beam splitter and two-mode squeezing operations favors the efficient charging and promotes the generation of ergotropy. Furthermore, charging performance can be enhanced by the squared electromagnetic vector potential $\mathbf{A}^2$ term in the deep strong-coupling regime. This work provides insights into the operational mechanisms of quantum batteries within the ultrastrong coupling regime.


This work is supported by the China Postdoctoral Science Foundation under Grant No. 2024M760829,
the National Natural Science Foundation of China under Grants No. 12575009 and No. 12205193, and the
Jiangxi Natural Science Foundation, under Grant No. 20242BAB20035.


The supporting information is available online, which includes Refs. \cite{PhysRevB.98.205423, PhysRevB.72.115303, PhysRevA.91.063840, PhysRevLett.107.100401, RevModPhys.91.025005, PhysRevB.99.035421}. The supporting materials are published as submitted, without typesetting or editing. The responsibility for scientific accuracy and content remains entirely with the authors.

\bibliography{ref.bib}

\clearpage

\setcounter{equation}{0}
\setcounter{page}{1}
\setcounter{figure}{0}
\renewcommand{\thepage}{S\arabic{page}}  
\renewcommand{\thefigure}{S\arabic{figure}}
\renewcommand{\theequation}{S\arabic{equation}}
\onecolumngrid

\renewcommand{\thepage}{S\arabic{page}}  
\renewcommand{\thefigure}{S\arabic{figure}}
\renewcommand{\theequation}{S\arabic{equation}}

\appendix

\begin{center}
\textbf{Supplemental Material:}\\
\textbf{Dissipative Quantum Battery in the Ultrastrong Coupling Regime Between Two Oscillators}

\end{center}

\section{The diagonalization of the system Hamiltonian} \label{Appendix A}

The eigenvalues and eigenstates of the system Hamiltonian Eq. (1) in the main text are determined by exploiting the commutation relation $\left[ A_{j}, H_{S}\right]=\omega_{j} A_{j}$, which leads to the following eigenvalue equation,
\begin{equation} \tag{S1} \label{eig-Eq}
\left(\begin{array}{cccc}
\omega_{a} & g & 0 & g \\
g & \omega_{b} & g & 0 \\
0 & -g & -\omega_{a} & -g \\
-g & 0 & -g & -\omega_{b}
\end{array}\right)
\left(\begin{array}{cccc}
t_{j}  \\
u_{j} &   \\
v_{j} & \\
w_{j}
\end{array}\right)=\omega_{j}\left(\begin{array}{cccc}
t_{j}  \\
u_{j} &   \\
v_{j} & \\
w_{j}
\end{array}\right).
\end{equation}
The corresponding unnormalized Hopfield coefficients are obtained by solving this equation as 
$t_{\pm}=\mp \frac{\omega_c \pm (\omega_a+\omega_b)(\omega_a+\omega_b+2 \omega_{\pm})}{4 g \omega_a}$, $u_{\pm}=-1+\frac{2 \omega_b}{\omega_b-\omega_{\pm}}$, $v_{\pm}=\frac{\pm \omega_c+(\omega_a-\omega_b)(\omega_a-\omega_b-2 \omega_{\pm})}{4 g \omega_a}$, and $w_{\pm}=1$ with $\omega_c$ and $\omega_{\pm}$ defined as before. In the resonant case, where $\omega_a=\omega_{b}=\omega$, the normal frequencies are given $\omega_{\pm}=\sqrt{\omega(\pm 2 g +\omega)}$, and the related Hopfield coefficients are $t_{\pm}=-\frac{\pm g+\omega+\omega_{\pm}}{\mathcal{N}_{\pm} g}$, $u_{\pm}=-\frac{\pm g+\omega+\omega_{\pm}}{\mathcal{N}_{\pm} g}$, $v_{\pm}=\pm\frac{1}{\mathcal{N}_{\pm}}$ and $w_{\pm}=\frac{1}{\mathcal{N}_{\pm}}$.

When considering the interaction of a beam-splitter or two-mode squeezing, as shown in Eq. (25), the eigenvalue equation takes the following form,
\begin{equation} \tag{S2} \label{eig-Eq1}
\left(\begin{array}{cccc}
\omega_{a} & g_{\mathrm{bs}} & 0 & g_{\mathrm{sq}} \\
g_{\mathrm{bs}} & \omega_{b} & g_{\mathrm{sq}} & 0 \\
0 & -g_{\mathrm{sq}} & -\omega_{a} & -g_{\mathrm{bs}} \\
-g_{\mathrm{sq}}& 0 & -g_{\mathrm{bs}} & -\omega_{b}
\end{array}\right)
\left(\begin{array}{cccc}
t_{j}  \\
u_{j} &   \\
v_{j} & \\
w_{j}
\end{array}\right)=\omega_{j}\left(\begin{array}{cccc}
t_{j}  \\
u_{j} &   \\
v_{j} & \\
w_{j}
\end{array}\right).
\end{equation}
When $g_{\mathrm{sq}}=0$, the normal frequencies can be solved as $\omega_{\pm}=\frac{\omega_a \pm \sqrt{4 g_{\mathrm{bs}} +(\omega_a-\omega_b)^2} +\omega_b}{2}$, and the Hopfield coefficients are $t_{\pm}=\frac{\omega_{\pm}-\omega_b}{g_{\mathrm{bs}}}$, and $u_{\pm}=1$. In the case of $g_{\mathrm{bs}}=0$, the two positive frequencies are given by $\omega_{1, 2}=\frac{\pm(\omega_a-\omega_b)+\sqrt{-4 g_{\mathrm{sq}}^2+(\omega_a+\omega_b)^2}}{2}$. The corresponding unnormalized Hopfield coefficients are $t_{1}=-\frac{ \omega_{1}+\omega_b}{ g_{\mathrm{sq}}}$, $w_{1}=1$, $u_{2}=-\frac{\omega_2+\omega_a}{ g_{\mathrm{sq}}}$, and $v_2=1$. 
This formulation provides a clear description of the eigenvalue problem in the two limiting cases: either only a beam splitter or only two-mode squeezing.

In addition, as shown in Eq. (28), in order to evaluate the ergotropy, we can calculate $\mathcal{M}$ as $\mathcal{M}= (1+2 (1-\mathrm{e}^{-\gamma^{a}_{-} \frac{\sin^2 \theta \omega_{a}}{\omega_{-}} t}) N^{a}(\omega_-))^2 \cos^4 \theta+(1+2 (1-\mathrm{e}^{-\gamma^{a}_{+} \frac{\cos^2 \theta \omega_{a}}{\omega_{+}} t}) N^{a}(\omega_+))^2 \sin^4 \theta+[(1+2 (1-\mathrm{e}^{-\gamma^{a}_{-} \frac{\sin^2 \theta \omega_{a}}{\omega_{-}} t}) N^{a}(\omega_-))(1+2 (1-\mathrm{e}^{-\gamma^{a}_{+} \frac{\cos^2 \theta \omega_{a}}{\omega_{+}} t}) N^{a}(\omega_+)) \cos^2 \theta \sin^2 \theta (\omega_{-}^2+ \omega_{+}^2)]/(\omega_{-} \omega_{+})$. 

\section{The solution of the second moments of normal operators } \label{Appendix B}

With Eq. (11), the dynamics of second moments for the system in the normal basis can be derived in the following form 
\begin{align}\tag{S3} \label{Eq:second_moments}
\frac{\mathrm{d} \Phi}{\mathrm{d} t}=M \Phi +Q,
\end{align}
where 
$\Phi=(\langle A_{+}^2  \rangle, \langle A_{-}^2  \rangle, \langle  A_{+} A_{-} \rangle, \langle A_{+}^{\dagger} A_{+} \rangle, \langle A_{-}^{\dagger} A_{-} \rangle, \langle A_{+}^{\dagger} A_{-} \rangle)^{\mathrm{T}}$ and 
$Q=[0, 0, 0, \Gamma_{a}(\omega_{+})|W_{+}|^2$, $\Gamma_{a}(\omega_{-})|W_{-}|^2, 0]^{\mathrm{T}}$.
The dynamical matrix $M$ is diagonal, meaning that only its principal diagonal contains nonzero elements. The diagonal entries are given by
\begin{align} \tag{S4}
\begin{split}
M_{1,1}&=-2 \mathrm{i} \omega_{+}-\gamma^a_{+} |W_{+}|^2,\\
M_{2,2}&=-2 \mathrm{i} \omega_{-}-\gamma^a_{-} |W_{-}|^2,\\
M_{3,3}&=-\sum_{j}\left\lbrace \frac{\gamma^a_{j} |W_{j}|^2}{2}+\mathrm{i}\omega_j \right\rbrace, \\
M_{4,4}&=-\gamma^a_{+} |W_{+}|^2,\\
M_{5,5}&=-\gamma^a_{-} |W_{-}|^2, \\
M_{6,6}&=\mathrm{i} (\omega_{+}-\omega_{-})-\sum_{j}\frac{\gamma^a_{j} |W_{j}|^2}{2}. \\
\end{split}
\end{align}
The remaining second moments may be derived from these. The dynamics of these second moments can be solved with initial conditions as
\begin{align} \tag{S5}
\begin{split}
\langle A_{j}^2\rangle&= \mathrm{e}^{-t \gamma^{a}_{j} |W_{j}|^2 -2 \mathrm{i} t \omega_{j}} \langle A_{j}^2\rangle|_{t=0}, \\
\langle A_{j}^{\dagger} A_{j} \rangle&=\mathrm{e}^{-t \gamma^{a}_{j} |W_{j}|^2} [\langle A_{j}^{\dagger} A_{j} \rangle |_{t=0}+ N(\omega_{j})(-1+\mathrm{e}^{t \gamma^{a}_{j} |W_{j}|^2})],\\ 
\langle A_{+}^{\dagger} A_{-}\rangle &= \mathrm{e}^{-\sum_{j}\frac{t \gamma^{a}_{j} |W_{j}|^2}{2}+\mathrm{i} (\omega_{+}-\omega_{-}) t} \langle A_{+}^{\dagger} A_{-}\rangle |_{t=0}, \\
   \langle A_{-} A_{+} \rangle &= \mathrm{e}^{-\sum_{j}(\frac{ t \gamma^{a}_{j} |W_{j}|^2}{2}+\mathrm{i} \omega_{j} t)} \langle A_{-} A_{+} \rangle |_{t=0}.
   \end{split}
\end{align}

For initial state $\rho(0)=|00 \rangle_{ab} \langle 00|$, the initial second correlation quantities as follows 
\begin{align} \tag{S6} \label{eq:initial_condition_bare}
\begin{split}
\langle A_{j}^2 \rangle|_{t=0}& =t_{j} v_{j}+u_{j} w_{j},  \\ 
\langle A_{j}^{\dagger} A_{j}\rangle |_{t=0}&=|v_{j}|^2+|w_{j}|^2,\\ 
 \langle A_{+}^{\dagger} A_{-} \rangle |_{t=0}&=v_{+}^{*} v_{-}+w_{+}^{*} w_{-},\\
\langle A_{-} A_{+}\rangle |_{t=0}&=t_{-} v_{+}+u_{-} w_{+}. 
\end{split}
\end{align} 
The second moments can be solved as 
\begin{align} \tag{S7} \label{eq:second_moments_bare}
\begin{split}
\langle A_{j}^2\rangle&= \mathrm{e}^{-t \gamma^{a}_{j} |W_{j}|^2 -2 \mathrm{i} t \omega_{j}} (t_{j} v_{j}+u_{j} w_{j}), \\
\langle A_{j}^{\dagger} A_{j} \rangle&=\mathrm{e}^{-t \gamma^{a}_{j} |W_{j}|^2} ((|v_{j}|^2+|w_{j}|^2)-N(\omega_{j})+\mathrm{e}^{t \gamma^{a}_{j} |W_{j}|^2} N(\omega_{j})),\\ 
 \langle A_{+}^{\dagger} A_{-}\rangle&= \mathrm{e}^{-\frac{t \gamma^{a}_{-} |W_{-}|^2}{2}-\frac{t \gamma^{a}_{+} |W_{+}|^2}{2}-\mathrm{i} \omega_{-} t+\mathrm{i} \omega_{+} t} (v_{+}^{*} v_{-}+w_{+}^{*} w_{-}), \\
 \langle A_{-} A_{+} \rangle&= \mathrm{e}^{-\frac{t \gamma^{a}_{-} |W_{-}|^2}{2}-\frac{t \gamma^{a}_{+} |W_{+}|^2}{2}-\mathrm{i} \omega_{-} t-\mathrm{i} \omega_{+} t} (t_{-} v_{+}+u_{-} w_{+}).
 \end{split}
\end{align}
 Hence, one can determine the energy $E_{b}$ and ergotropy $\mathcal{E}$ of the quantum battery using Eqs. (5), (12) and (14) in the main text.

\section{The temporal evolutions of the field operators} \label{Appendix C}

Here, we present an analysis of energy storage in a closed system comprising two coupled bosonic modes. Following the procedure outlined in Ref. \cite{PhysRevB.98.205423}, the time-dependent operators $a(t)$ and $b(t)$ can be determined by solving the following equations with initial conditions $a(0)$ and $b(0)$,
\begin{align} \tag{S8}
\left(\begin{array}{c}
a(t)   \\
b(t)  \\
a^{\dagger}(t)   \\
b^{\dagger}(t) 
\end{array}\right)=\left(\begin{array}{cccc}
M_{11} & M_{12} & M_{13} & M_{14}  \\
M_{21} & M_{22} & M_{23} & M_{24}  \\
M_{31} & M_{32} & M_{33} & M_{34}  \\
M_{41} & M_{42} & M_{43} & M_{44}  
\end{array}\right)
\left(\begin{array}{c}
a(0)   \\
b(0) \\
a^{\dagger}(0)   \\
b^{\dagger}(0) 
\end{array}\right).
\end{align}

\begin{figure}[H]
 \centering
\includegraphics[width=0.3\columnwidth]{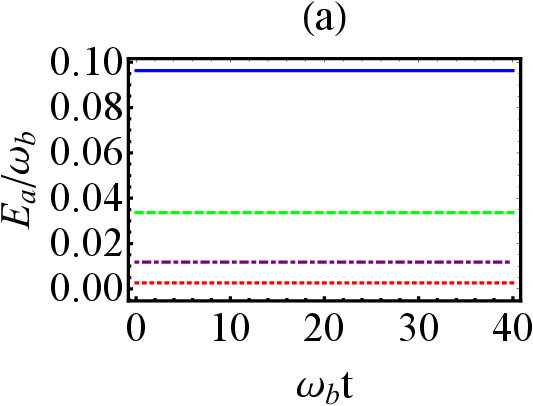}
\includegraphics[width=0.3\columnwidth]{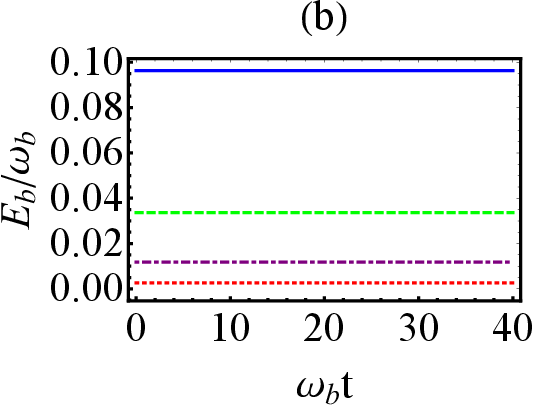}
\includegraphics[width=0.3\columnwidth]{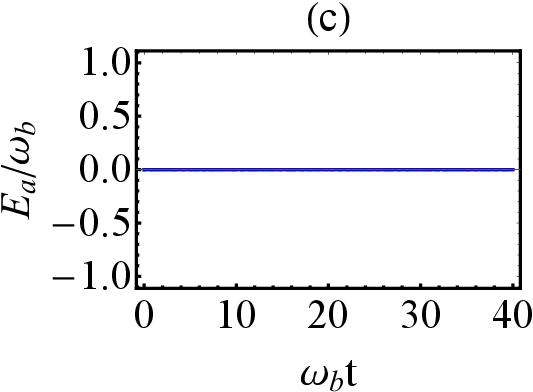}
\includegraphics[width=0.3\columnwidth]{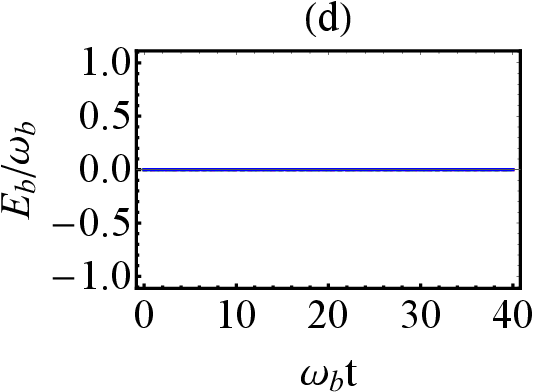}
\includegraphics[width=0.3\columnwidth]{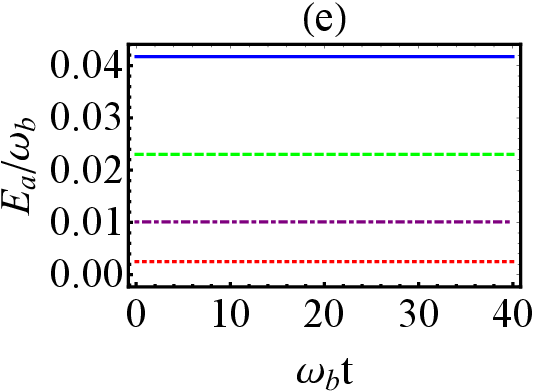}
\includegraphics[width=0.3\columnwidth]{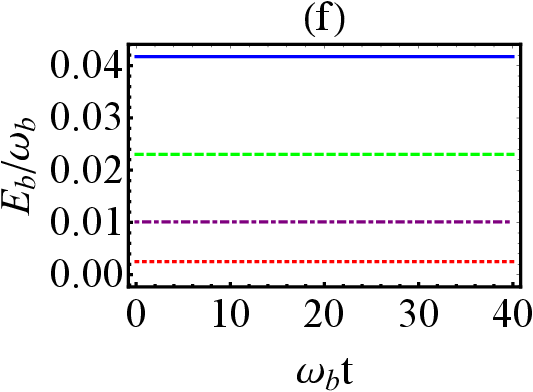}
\caption{\label{Fig:close_polariton}
The energy $E_{a}$ (a, c, e) and $E_{b}$ (b, d, f) versus the time $t$ for different coupling strengths $g$. The panels (a, b), (c, d), and (e, f) depict the full interaction, the beam-splitter interaction, and the two-mode squeezing interaction, respectively. Moreover, the dotted red, dash-dotted purple, dashed green, and solid blue lines correspond to coupling strengths of $g/\omega_b=0.1$, $g/\omega_b=0.2$, $g/\omega_b=0.3$, and $g/\omega_b=0.4$, respectively. All other parameters are set such that $\omega_{a}=\omega_{b}=\omega$, and all values are expressed in units of the frequency $\omega_b$. }
\end{figure}

For the initial state $| G \rangle=|0 \rangle_{+} |0 \rangle_{-}$, this state is the ground state of the system in the ultrastrong coupling regime, and it contains finite excitations \cite{PhysRevB.72.115303}. The energy of the charger and battery for the resonant case can be expressed as $E_a=E_b=\sum_{j}\frac{\omega}{2 (\omega+\omega_{j})^2 \omega_j}$. As shown in Figs. \ref{Fig:close_polariton}(a) and \ref{Fig:close_polariton}(b), the energies of the quantum charger and battery remain unchanged because the state considered is the ground state of the system, i.e., $A_{j} | 0 \rangle_{j}=0$. Moreover, from Fig. \ref{Fig:close_polariton} (c), Fig. \ref{Fig:close_polariton} (d), Fig. \ref{Fig:close_polariton} (e), and Fig. \ref{Fig:close_polariton} (f), parametric interaction facilitates energy storage, whereas beam-splitter interaction does not.

\begin{figure}[H]
\centering
\includegraphics[width=0.3\columnwidth]{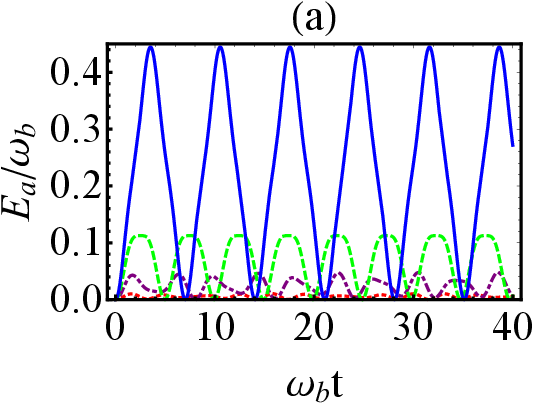}
\includegraphics[width=0.3\columnwidth]{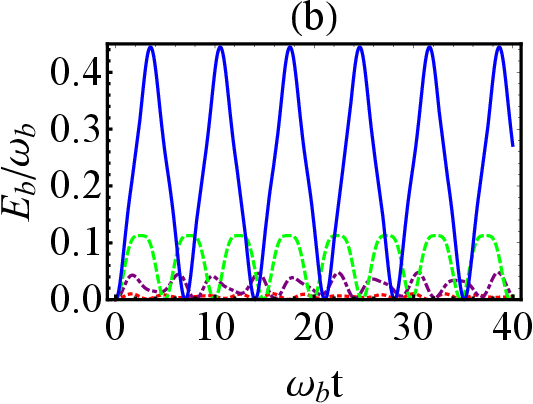}
\includegraphics[width=0.3\columnwidth]{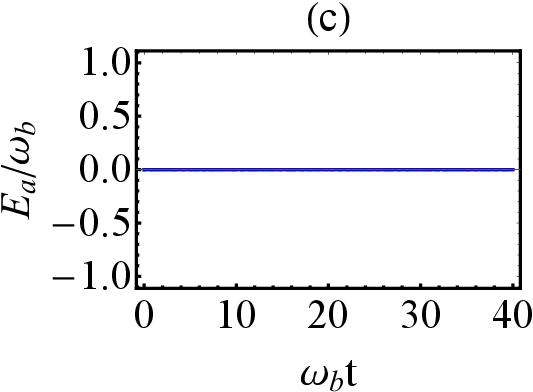}
\includegraphics[width=0.3\columnwidth]{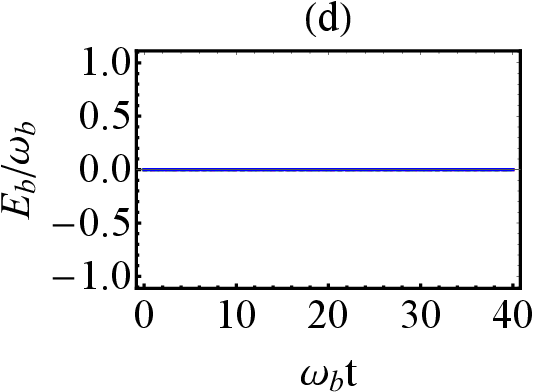}
\includegraphics[width=0.3\columnwidth]{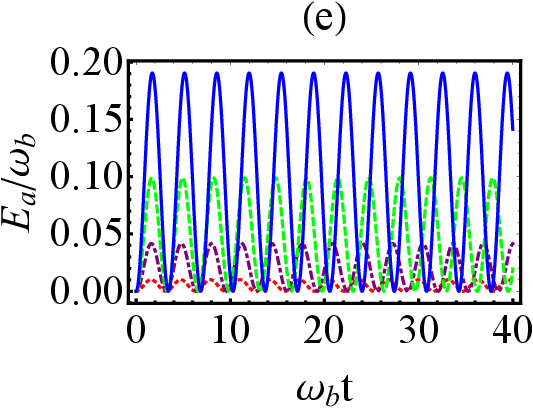}
\includegraphics[width=0.3\columnwidth]{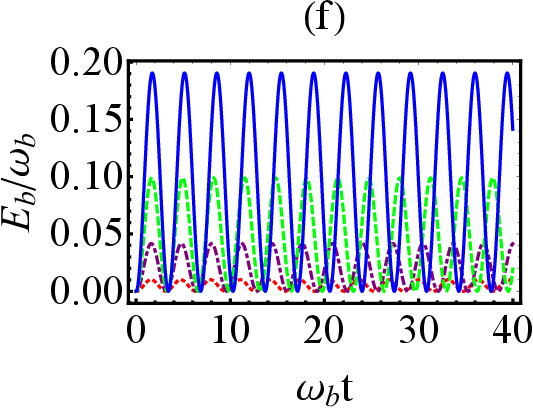}
\caption{\label{fig:qb_close_bare}
The energy $E_{a}$ (a, c, e) and $E_{b}$ (b, d, f) versus the time $t$ for different coupling strength $g$. The panels (a, b), (c, d), and (e, f) depict the full interaction, the beam-splitter interaction, and the two-mode squeezing interaction, respectively. Moreover, the dotted red, dashed-dotted purple, dashed green, and solid blue lines represent the coupling strengths $g/\omega_b=0.1$, $g/\omega_b=0.2$, $g/\omega_b=0.3$, and $g/\omega_b=0.4$, respectively. The other parameters are set as $\omega_{a}=\omega_{b}=\omega$, with all values expressed in units of the frequency $\omega_b$.}
\end{figure}

For the initial state $\rho(0)=|00 \rangle_{ab} \langle 00|$, the system no longer resides in its ground state in the ultrastrong coupling regime \cite{PhysRevA.91.063840}. Accordingly, the energies can be calculated as
\begin{align} \tag{S9}
E_{a}&=\omega (|M_{13}|^2+|M_{14}|^2),\\
E_{b}&=\omega (|M_{23}|^2+|M_{24}|^2), \tag{S10}
\end{align}
where $M_{13}=\frac{\mathrm{i} g(g-\omega-\omega_{-}) \sin (\omega_{-} t)}{(\omega+\omega_{-})(-2 g+\omega+\omega_{-})}+\frac{\mathrm{i} g (g+\omega+\omega_{+}) \sin (\omega_{+} t)}{(\omega+\omega_{+})(2 g+\omega+\omega_{+})}$ and $M_{14}=-\frac{\mathrm{i} g(g-\omega-\omega_{-}) \sin (\omega_{-} t)}{(\omega+\omega_{-})(-2 g+\omega+\omega_{-})}+\frac{\mathrm{i} g (g+\omega+\omega_{+}) \sin (\omega_{+} t)}{(\omega+\omega_{+})(2 g+\omega+\omega_{+})}$. Because $M_{23}=M_{14}$, and $M_{24}=M_{13}$, the condition $E_{b}=E_{a}$ is naturally satisfied for resonant case. As depicted in Fig. \ref{fig:qb_close_bare}, the energies of the charger and battery exhibit a Rabi-like oscillatory behavior, and both the beam-splitter and two-mode squeezing interactions enable greater energy storage than other cases, consistent with the observations in Fig. \ref{Fig:close_polariton}.

\section{Compared to other battery proposals} \label{Appendix D}
\begin{figure}[H]
\centering
\includegraphics[width=0.6\columnwidth]{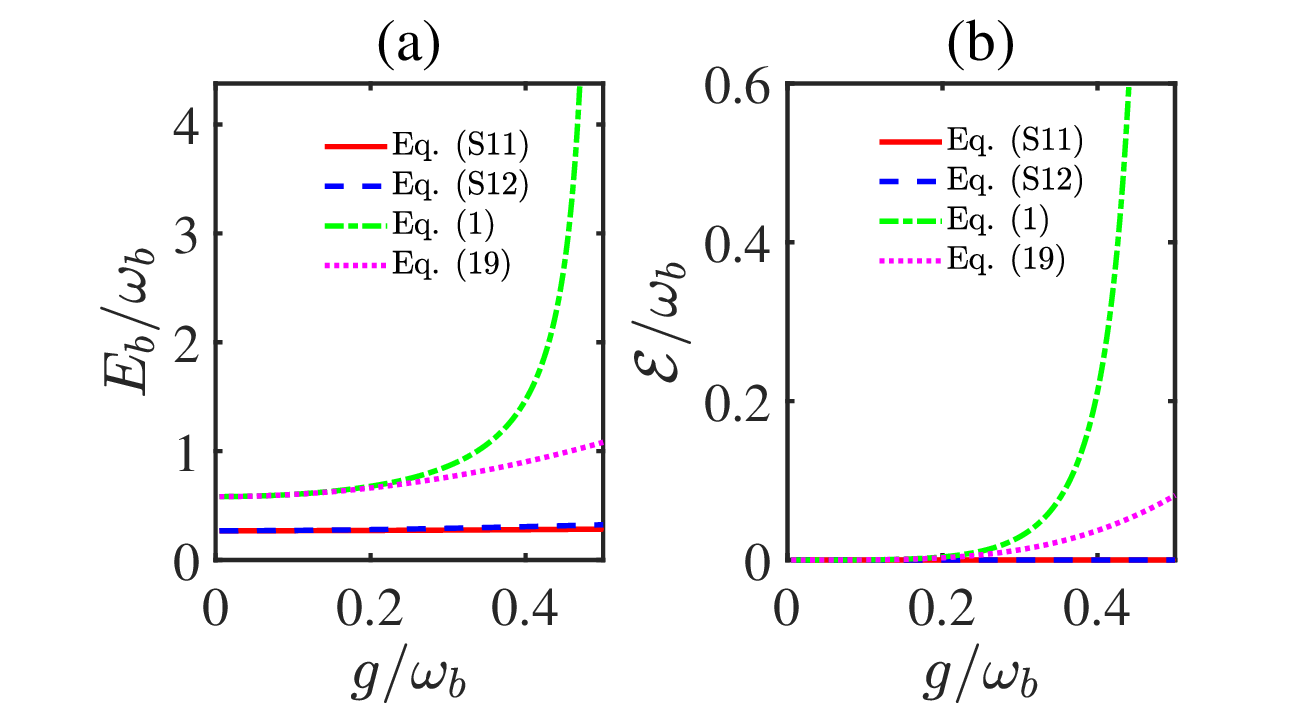}
\caption{\label{fig:qb_com}
The steady-state energy $E_{b}$ (a) and its corresponding ergotropy $\mathcal{E}$ (b) versus the coupling strength $g$. The other parameters are set as $\omega_{a}=\omega_{b}=\omega$, $\gamma^{a}=10^{-3} \omega_b$ and $T_{a}=\omega_{b}$, with all values expressed in units of the frequency $\omega_b$.}
\end{figure}

Here, we further consider the performance of the battery using the two-qubit and quantum Rabi models.
 The Hamiltonian for two coupled qubits is given by
\begin{align} \tag{S11}\label{H_S1}
H_S=\frac{\omega_{a} (\sigma^{z}_{a}+\mathbb{1}_{2})}{2}+\frac{\omega_{b}(\mathbb{1}_{2}+ \sigma_{b}^{z})}{2}+g \sigma_{a}^{x} \sigma_{b}^{x},
\end{align}
and the interaction Hamiltonian between qubit $a$ and the heat reservoir is expressed as
$H_{R-a}=\sum_k g_{k}[(\sigma_{a}^{+}+\sigma_{a}^{-})(d^{\dagger}_{k} +d_{k})]$.
In the quantum Rabi model, the system Hamiltonian can be written as \cite{PhysRevLett.107.100401, RevModPhys.91.025005}
\begin{align} \tag{S12}\label{H_S2}
H_S=\omega_{a} a^{\dagger} a+\frac{\omega_{b}(\mathbb{1}_{2}+\sigma_{b}^{z})}{2}+g (a^{\dagger}+a) \sigma_{b}^{x},
\end{align}
and the system–environment coupling Hamiltonian is
$H_{R-a}=\sum_k g_{k}[(a^{\dagger}+a)(d^{\dagger}_{k} +d_{k})]$. For a two-level atom quantum battery, its ergotropy can be solved by \cite{PhysRevB.99.035421}
\begin{align} \tag{S13}
\mathcal{E}=\frac{\omega_b}{2}(\sqrt{\langle \sigma^{z}_{b} \rangle^2+4 \langle \sigma^{+}_{b} \rangle \langle \sigma^{-}_{b} \rangle}+\langle \sigma^{z}_{b} \rangle).
\end{align}
We employ the same procedure to analyze the energy storage and ergotropy of quantum batteries across these different models. As illustrated in Fig. \ref{fig:qb_com}, the two-oscillator battery outperforms the other models in the ultrastrong coupling regime. Notably, no ergotropy is generated in either the quantum two-qubit or the Rabi models.

\section{ The mechanism of unidirectional energy transfer for weakly coupled oscillators with local master equation } \label{Appendix E}

Here we would like to analyze the battery for a weakly coupled oscillator system shown in Ref. \cite{PhysRevB.99.035421}. The local master equation for dissipative oscillators in the weak internal coupling regime is 
\begin{align} \tag{S14}
\frac{\mathrm{d} \rho}{\mathrm{d} t}&=-\mathrm{i}[H_S, \rho]+\frac{\gamma N(\omega)}{2}\mathcal{D}(a^{\dagger}, a) \rho+\frac{\gamma (N(\omega)+1)}{2}\mathcal{D}(a, a^{\dagger}),
\end{align}
where $H_S=\omega a^{\dagger} a+\omega b^{\dagger} b +g(a b^{\dagger}+a^{\dagger} b)$, and the dissipator is $\mathcal{D}(o, o^{\prime})\rho=o \rho o^{\prime \dagger}-\frac{\lbrace o^{\prime \dagger} o, \rho \rbrace}{2}$.
The local master equation in the global representation with $A_{\pm}=\frac{a \pm a^{\dagger}}{\sqrt{2}}$ can be written as
\begin{align} \tag{S15}
\begin{split}
\frac{\mathrm{d} \rho}{\mathrm{d} t}&=-\mathrm{i}[\omega_{+} A_{+}^{\dagger} A_{+}+\omega_{-} A_{-}^{\dagger} A_{-}, \rho]+\frac{\gamma N(\omega)}{2}[\mathcal{D}(A_{+}^{\dagger}, A_{+}) \rho +\mathcal{D}(A_{+}^{\dagger}, A_{-}) \rho +\mathcal{D}(A_{-}^{\dagger}, A_{+}) \rho+\mathcal{D}(A_{-}^{\dagger}, A_{-}) \rho] \\&+\frac{\gamma N(-\omega)}{2}[\mathcal{D}(A_{+}, A_{+}^{\dagger}) \rho +\mathcal{D}(A_{+}, A_{-}^{\dagger}) \rho+ \mathcal{D}(A_{-}, A_{+}^{\dagger}) \rho +\mathcal{D}(A_{-}, A_{-}^{\dagger}) \rho].
\end{split}
\end{align}
The dynamical equations of the second moments are 
\begin{align}\tag{S16}\label{eq_second_moments_lme}
\begin{split}
\frac{\mathrm{d} \langle A_{j}^{\dagger} A_{j} \rangle}{\mathrm{d} t}&=\frac{\gamma}{2}(N(\omega)- \langle A^{\dagger}_{j} A_{j} \rangle)-\frac{\gamma}{2} (\langle A_{-} A_{+}^{\dagger} \rangle+\langle A_{-}^{\dagger} A_{+}\rangle),\\
\frac{\mathrm{d} \langle A_{j}^2\rangle}{\mathrm{d} t}&=-(2 \mathrm{i} \omega_{j}+\frac{\gamma}{2})\langle A_{j}^2 \rangle-\frac{\gamma }{2} \langle A_{+} A_{-} \rangle,\\
\frac{\mathrm{d} \langle A_{-} A_{+}^{\dagger} \rangle}{\mathrm{d} t}&=
(2 \mathrm{i} g-\frac{\gamma}{2}) \langle A_{-} A_{+}^{\dagger} \rangle+\frac{\gamma}{2}(N(\omega)-\frac{\langle A_{+}^{\dagger} A_{+}\rangle+\langle A_{-}^{\dagger} A_{-}\rangle}{2}),\\
\frac{\mathrm{d} \langle A_{+} A_{-}\rangle}{\mathrm{d} t}&=-[\mathrm{i}(\omega_{+}+\omega_{-})+\frac{\gamma }{2}] \langle A_{+} A_{-}\rangle-\frac{\gamma }{4} (\langle A_{+}^2\rangle+\langle A_{-}^2\rangle).
\end{split}
\end{align}
The terms including $\langle A_{\pm}^{\dagger} A_{\pm}\rangle$, $\langle A_j^2 \rangle$, $\langle A_{+} A_{-}\rangle$ and $\langle A_{-} A_{+}^{\dagger} \rangle$ exhibit mutual dependence, which represents a key difference from the behavior observed in systems with two ultrastrongly coupled oscillators.
In this case, the ground state of the system is $|G\rangle=|00\rangle_{ab}$. So, the nonzero second moments can be solved as 
\begin{align} \tag{S17}
\langle A_{\pm}^{\dagger} A_{\pm} \rangle&=N(\omega)\lbrace 1-\frac{\mathrm{e}^{-\gamma t/2}}{G^2}(16 g^2-\gamma^2 \cos(G t/2)) \rbrace,\\
\langle A_{-} A_{+}^{\dagger} \rangle&=N(\omega)\lbrace \frac{\mathrm{e}^{-\gamma t/2}}{G} \gamma \sin(\frac{G t}{2})+\mathrm{i} \frac{4  \mathrm{e}^{-\gamma t/2}}{G^2}  \gamma g(1-\cos(\frac{G t}{2}))\rbrace, \tag{S18}
\end{align}
where $G=\sqrt{(4 g)^2-\gamma^2}$.
Based on the relations $A_{\pm}=\frac{a \pm a^{\dagger}}{\sqrt{2}}$, the energy of battery $E_b=\omega \langle b^{\dagger} b\rangle$ can be solved for two regimes, respectively as
\begin{equation} \tag{S19}
\frac{E_b}{\omega}= \left \lbrace
\begin{aligned}
& N(\omega) \lbrace 1+\frac{\mathrm{e}^{-\frac{\gamma \tau}{2}}}{\Gamma^2}(16 g^2-\gamma \Gamma \sinh(\frac{\Gamma t}{2})-\gamma^2 \cosh(\frac{\Gamma t}{2})) \rbrace, \quad \gamma > 4 g, \\
& N(\omega) \lbrace 1-\frac{\mathrm{e}^{-\frac{\gamma t}{2}}}{G^2}(16 g^2+\gamma G \sin(\frac{G t}{2})-\gamma^2 \cos(\frac{G t}{2})) \rbrace, \quad \gamma<4 g ,
\end{aligned}
\right.
\end{equation}
where $\Gamma=\sqrt{\gamma^2-(4 g)^2}$.
Hence, it can exhibit monotonically increasing energy transfer in the overdamped regime $\gamma>4 g$, although it has no ergotropy.





\end{document}